\documentclass[10pt]{article}
\usepackage{amsmath}
\usepackage{amssymb}
\usepackage{enumerate}
\usepackage{psfrag}
\usepackage{mathrsfs}
\usepackage{hyperref}
\usepackage{slashed}
\usepackage{mathrsfs}
\usepackage{amsthm} 
\newtheorem{innercustomthm}{Theorem}
\newenvironment{customthm}[1]
  {\renewcommand\theinnercustomthm{#1}\innercustomthm}
  {\endinnercustomthm}
\usepackage[dvips]{graphicx}

\DeclareGraphicsExtensions{.eps,.art,.ART,.ps}
\newcommand{\bbR}{\mathbb{R}}      

\newcommand{\vertiii}[1]{{\left\vert\kern-0.2ex\left\vert\kern-0.2ex\left\vert #1 \right\vert\kern-0.2ex\right\vert\kern-0.2ex\right\vert}}
\newtheorem{Thm}{Theorem}[section]

\theoremstyle{definition}
\newtheorem{Def}[Thm]{Definition}

\begin{document}
\title{Solutions of the wave equation bounded at the Big Bang}
\author{Pedro Gir\~{a}o, Jos\'e Nat\'ario and Jorge Silva\\ \\
{\small CAMGSD, Departamento de Matem\'{a}tica, Instituto Superior T\'{e}cnico,}\\
{\small Universidade de Lisboa, Portugal}
}
\date{}
\maketitle
\begin{abstract}
By solving a singular initial value problem, we prove the existence of solutions of the wave equation $\Box_g\phi=0$ which are bounded at the Big Bang in the Friedmann-Lema\^\i tre-Robertson-Walker cosmological models. More precisely, we show that given any function $A \in H^3(\Sigma)$ (where $\Sigma=\bbR^n, \mathbb{S}^n$ or $\mathbb{H}^n$ models the spatial hypersurfaces) there exists a unique solution $\phi$ of the wave equation converging to $A$ in $H^1(\Sigma)$ at the Big Bang, and whose time derivative is suitably controlled in $L^2(\Sigma)$.
\end{abstract}
\tableofcontents
%
%
%
\section{Introduction and statement of the main result} \label{section0}
It was recently shown in~\cite{AlhoFranzen}, by a clever use of the vector field method, that solutions of the wave equation in the flat Friedmann-Lema\^\i tre-Robertson-Walker (FLRW) cosmological models, and also in the Kasner spacetime, generically blow up at the Big Bang singularity. Interestingly, however, not all (nonconstant) solutions blow up, and in fact one of the difficulties in~\cite{AlhoFranzen} is to formulate a genericity condition that excludes these special bounded solutions. This scenario, already found in \cite{Ringstrom}, also occurs for solutions of the wave equation in the black hole region of the Schwarzschild spacetime (which can be thought of as a cosmological model approaching a Big Crunch), as discussed in~\cite{GregSbierski}. The results in~\cite{Petersen2} suggest that a similar situation may occur near compact Cauchy horizons.

The purpose of this small note is to prove the existence of a large class of non-generic solutions of the wave equation bounded at the Big Bang, which had to be excluded in~\cite{AlhoFranzen}. Besides being interesting in themselves, these solutions are important in certain speculative cosmological scenarios, such as Conformal Cyclic Cosmology (described for instance in~\cite{Tod}), where one is concerned with solutions of the wave equation that propagate from one ``aeon" to the next, and are therefore bounded at the Big Bang (see \cite{Tod2} and references therein). They might also be relevant in the study of fields propagating  across the singularity in singular bouncing cosmologies (see for instance \cite{BattefeldPeter}).

Solutions of this kind have been previously found in the literature, although not in a systematic manner: they were often obtained as by-products of works with a different focus, and mostly correspond to particular cases of ours. A first example was provided by Klainerman and Sarnak \cite{KlainermanSarnak}, who gave the explicit solution for the wave equation for two particular FLRW models (the so-called Einstein-de Sitter universe and its hyperbolic analogue): their work was used in~\cite{AbbasiCraig} to prove the existence of solutions bounded at the Big Bang in these two models. The problem of prescribing asymptotic data at the Big Bang for the Einstein-de Sitter universe was further analyzed in~\cite{GalstianKinoshitaYagdjian, GalstianYagdjian}. More recently, Ringstr\"{o}m \cite{Ringstrom} studied linear systems of wave equations on general cosmological backgrounds with convergent asymptotics, including the problem of imposing asymptotic data. In fact, his wide ranging results (essentially) include several instances of Theorem~\ref{mainthm1} below, for example models with perfect fluid matter satisfying $\frac23 + \frac4{3n} < \gamma \leq 2$ (see Appendix~\ref{appendixA}).\footnote{We thank Hans Ringstr\"{o}m for pointing this out.} In a subsequent paper \cite{Ringstrom2}, he also determined the asymptotic behavior of solutions of the Klein-Gordon equation towards the Big Bang of Bianchi backgrounds, although in this case his approach does not allow for prescribing this behavior. The same issue has been addressed in~\cite{Bachelot}, for the conformally invariant wave equation (and also for other choices of the conformal coupling parameter which do not include the standard wave equation). The much harder problem of studying the nonlinear stability of the Big Bang singularity for perturbations of FLRW solutions, where one has to deal with the full Einstein-scalar field system, was treated in \cite{RodnianskiSpeck1, RodnianskiSpeck2, Speck}.

Our main result is the following:\footnote{We denote by $L^2(\Sigma)$ the standard Hilbert space of square Lebesgue-integrable functions on $\Sigma$ (with respect to the usual Riemannian measure), and by $H^k(\Sigma)$ the Sobolev space of functions on $\Sigma$ whose weak partial derivatives up to order $k$ are in $L^2(\Sigma)$.}

\begin{customthm}{1} \label{mainthm1}

Let $(M,g)$ be an expanding $(n+1)$-dimensional FLRW model, that is, $M=\bbR^+ \times \Sigma$ and
$$
g = -dt^2 + a^2(t) h,
$$
where $(\Sigma,h)$ is a simply connected $n$-dimensional Riemannian manifold of constant curvature $0, 1$ or $-1$, and that the scaling factor, solving Friedmann's equations, satisfies $\dot{a}(t)>0$. Assume also that $\lim_{t \to 0^+}a(t)=0$ and $1/{a(t)}$ is integrable in a right neighborhood of the origin.\footnote{For example, $a(t) \sim t^p$ as $t \to 0^+$, with $0 < p < 1$; see Appendix~\ref{appendixA} for concrete examples. This is the condition for the Big Bang to be a silent singularity in Ringstr\"{o}m's terminology \cite{Ringstrom, Ringstrom2}.} Given $A\in H^3(\Sigma)$, there exists a unique solution of the wave equation 
$$
\Box_g \phi = 0
$$
in $C^0\left((0,T],H^1(\Sigma)\right)\cap C^1\left((0,T],L^2(\Sigma)\right)$ (for each $T>0$) such that
\begin{equation}
\lim_{t \to 0^+} \|\phi(t,\,\cdot\,) - A(\,\cdot\,)\|_{H^1(\Sigma)}=0 \label{dado-inicial-posicao_0}
\end{equation}
and
\begin{equation}
\lim_{t \to 0^+} \left( a(t)\,\|\partial_t\phi(t,\,\cdot\,)\|_{L^2(\Sigma)} \right)=0. \label{dado-inicial-velocidade_0}
\end{equation}
\end{customthm}

The proof of this theorem uses energy methods. As we shall see, the last condition in this theorem is slightly stronger than requiring the solution to have finite energy at the Big Bang. Although we are assuming an expanding FLRW model, we are only concerned with a neighborhood of the Big Bang, where this is always true; it is immediate to extend this result to re-collapsing models.

Notice that it is not clear whether Theorem~\ref{mainthm1} captures all solutions of the wave equation that do not blow up at the Big Bang (even if they are smooth or satisfy the slightly stronger assumption of having a pointwise smooth limit $A$ at $t = 0$), as they would also have to satisfy conditions \eqref{dado-inicial-posicao_0} and \eqref{dado-inicial-velocidade_0}. However, we do indeed suspect that this is the case for sufficiently regular solutions. Proving this would probably require the derivation of an accurate enough asymptotic expansion of all linear waves towards the singularity, as was done in \cite{GregSbierski}.

In many cosmological models, including inflation and bouncing models, the scalar field has a potential. If this potential is quadratic in the field, so that the wave equation is linear, then the results in \cite{Ringstrom, Ringstrom2} suggest that some version of Theorem~\ref{mainthm1} should still hold, perhaps requiring the potential to satisfy some additional conditions (which should include the usual Klein-Gordon equation). However, if the potential term in the wave equation is nonlinear then the qualitative behavior of the solution can be quite different, including the possible formation of singularities not related to the Big Bang.
%
%
\section{Proof of the main result}\label{section1}

In this section we present the proof of Theorem~\ref{mainthm1}.
Let $(\Sigma,h)$ be a simply connected Riemannian manifold of constant curvature $0, 1$ or $-1$, that is, $\bbR^n$, $\mathbb{S}^n$ or $\mathbb{H}^n$ with the standard metric. Consider an expanding FLRW model, given by $M = \bbR^+ \times \Sigma$ with the Lorentzian metric
$$
g = -dt^2 + a^2(t) h_{ij} dx^i dx^j ,
$$
where $\dot{a}(t)>0$ and the latin indices $i$ and $j$ run from $1$ to $n$. Defining the conformal time coordinate as
\begin{equation} \label{tau}
\tau = \int_{t_0}^t \frac{ds}{a(s)},
\end{equation}
for some $t_0>0$, the metric becomes
$$
g = a^2(\tau) \left(-d\tau^2 + h_{ij} dx^i dx^j \right).
$$
Note the abuse of notation $a(\tau)=a(t(\tau))$.
The wave equation in this background,
$$
\Box_g \phi = 0 \Leftrightarrow \partial_\mu \left(\sqrt{-g} \, \partial^\mu \phi \right) = 0 \Leftrightarrow \partial_\mu \left(a^{n+1} \sqrt{h} \, \partial^\mu \phi \right) = 0,
$$
can be written as
\begin{equation}\label{waveqntau}
- \partial_\tau \left( a^{n-1} \partial_\tau \phi \right) + a^{n-1} \Delta \phi = 0,
\end{equation}
where $\Delta$ is the Laplacian operator on $(\Sigma,h)$ and we assume for the time being that $\phi \in C^\infty(M)$. 

Recall that the energy-momentum tensor associated to the wave equation is
$$
T_{\mu\nu} = \partial_\mu \phi \, \partial_\nu \phi - \frac12 \partial_\alpha \phi \, \partial^\alpha \phi \, g_{\mu \nu},
$$
This energy-momentum tensor satisfies the Dominant Energy Condition (DEC), so that
$-T^{\mu\nu}X_\nu$ is causal and future pointing for each vector field $X$ which
is also causal and future pointing.
Choosing the multiplier vector field
$$
X = a^{1-n} \frac{\partial}{\partial\tau},
$$
we form the current
$$
J_{\mu} = T_{\mu\nu} X^{\nu},
$$
which we will use to derive energy inequalities. Let us introduce the following notation:

\begin{Def}
For $B \subset \Sigma$ a geodesic ball and $\tau_*>0$, we define
$B_{\tau_*}\subset\Sigma$ to be the geodesic ball that satisfies $\{\tau_*\}\times B_{\tau_*}=
D^+(\{0\}\times B)\cap\{\tau=\tau_*\}$.\footnote{Note that although $\tau>0$ on $M$ we can still define the future domain of dependence $D^+(\{0\}\times B)$ by using the conformal structure.}
\end{Def}

\begin{Def}
For $B \subset \Sigma$ a geodesic ball and $\tau>0$, we define the energy
\begin{eqnarray*}
E(\tau)\ =\
E(\tau,B)& =& \int_{\{\tau\} \times B_\tau} J_{\mu} N^{\mu}\,a^n dV_\Sigma\ =\ \int_{B_\tau} T_{00}\,
dV_\Sigma\\\
&=& \frac12\int_{B_\tau}  \left[ (\partial_\tau\phi)^2 + |\nabla\phi|^2 \right]\,dV_\Sigma,
\end{eqnarray*}
where $N =\frac{1}{a} \frac{\partial}{\partial\tau}$ is the future unit timelike normal and $dV_\Sigma$ is the volume element of $(\Sigma,h)$.
\end{Def}

The deformation tensor associated with the multiplier $X$ is
\begin{eqnarray*}
\Pi\ =\ \frac12 \mathcal{L}_Xg &=&   a^{2-n}a'\,(-d\tau^2+h_{ij} dx^i dx^j) +(n-1)a^{2-n}a'd\tau^2 \\
&=&(n-2)a^{2-n} a' d\tau^2+a^{2-n}a'h_{ij} dx^i dx^j,
\end{eqnarray*}
where the prime denotes differentiation with respect to $\tau$.
Noting that
$$
T^{00}=\frac{1}{2a^4}\left[(\partial_\tau\phi)^2 + |\nabla\phi|^2\right]
$$
and
$$
T^{ij}=\frac{1}{a^4}h^{ik}\partial_k\phi h^{jl}\partial_l\phi-\,\frac{1}{2a^4}
h^{ij}\left[-(\partial_\tau\phi)^2 + |\nabla\phi|^2\right],
$$
we have
\begin{eqnarray}
T^{\mu\nu}\nabla_\mu X_\nu\ =\ T^{\mu\nu} \Pi_{\mu\nu}& = & \frac{(n-2)}2  a^{-2-n} a'
\left[(\partial_\tau\phi)^2 +|\nabla\phi|^2\right] \nonumber \\ &&
 +a^{-2-n}a'|\nabla\phi|^2-\,\frac{n}{2}a^{-2-n}a'\left[-(\partial_\tau\phi)^2 +|\nabla\phi|^2\right]  \nonumber \\
&= & (n-1)  a^{-2-n}a' (\partial_\tau\phi)^2 \geq 0 \label{bulkpositive} 
\end{eqnarray}
(recall that we are assuming $a' > 0$).

\begin{figure}[h!]
\begin{center}
\psfrag{t=t0}{$\tau=\tau_0$}
\psfrag{t=t1}{$\tau=\tau_1$}
\psfrag{B0}{$\{\tau_0\}\times B_{\tau_0}$}
\psfrag{B1}{$\{\tau_1\}\times B_{\tau_1}$}
\psfrag{R}{$\mathcal{R}$}
\leavevmode
\includegraphics[scale=.4]{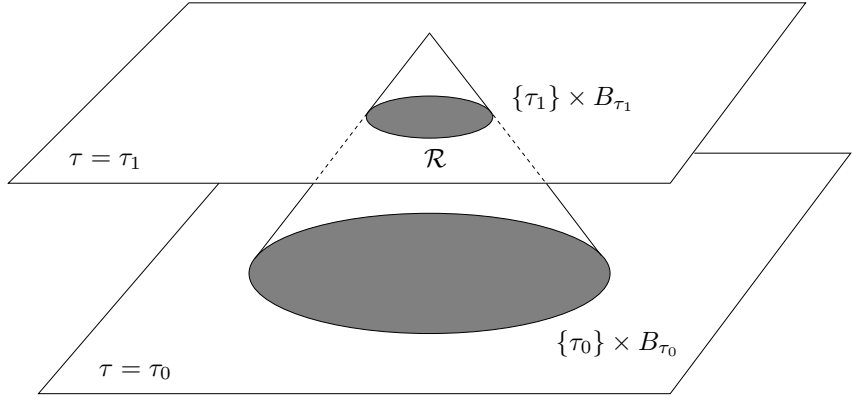}
\end{center}
\caption{Region $\mathcal{R}$ and its boundary.}\label{regionR}
\end{figure}

To obtain a priori estimates for the wave equation \eqref{waveqntau} we apply the divergence theorem to the current $J$ in the region
$$
\mathcal{R} =
\mathcal{R}(\tau_0,\tau_1, B) =
D^+(\{\tau_0\} \times B_{\tau_0}) \cap \{ \tau \leq \tau_1 \},
$$
where $\tau_0<\tau_1$ (see Figure~\eqref{regionR}):
$$
\int_\mathcal{R}\nabla_\mu J^\mu =
\int_\mathcal{R}(\nabla_\mu T^{\mu\nu})X_\nu+
\int_\mathcal{R} T^{\mu\nu}\nabla_\mu X_\nu
=\int_{\partial\mathcal{R}}
J_\mu N^\mu.
$$

Due to the DEC, the flux across the future null boundaries is nonpositive.
Moreover, for $\phi\in C^\infty(M)$ we have $(\nabla_\mu T^{\mu\nu})X_\nu=(\Box_g\phi)(X\cdot\phi)$. Therefore, when $\phi$ is a solution of the wave equation \eqref{waveqntau}, we obtain
\begin{equation}\label{divergence}
E(\tau_1)\leq E(\tau_0) \, ,
\end{equation}
that is,
\begin{equation}\label{energy-inequality_0}
\int_{B_{\tau_1}}  \left[ (\partial_\tau\phi)^2 + |\nabla\phi|^2 \right](\tau_1,\,\cdot\,)\,dV_\Sigma\leq
\int_{B_{\tau_0}}  \left[ (\partial_\tau\phi)^2 + |\nabla\phi|^2 \right](\tau_0,\,\cdot\,)\,dV_\Sigma.
\end{equation}
By the Fundamental Theorem of Calculus, we have, for $\tau \geq \tau_0$,
\begin{eqnarray}
\|\phi(\tau,\,\cdot\,)\|_{L^2(B_\tau)}&\leq&
\|\phi(\tau_0,\,\cdot\,)\|_{L^2(B_\tau)}\nonumber+\int_{\tau_0}^{\tau}
\|\partial_\tau\phi(s,\,\cdot\,)\|_{L^2(B_\tau)}\,ds\nonumber\\
&\leq&\|\phi(\tau_0,\,\cdot\,)\|_{L^2(B_{\tau_0})}+(\tau-\tau_0)\sqrt{2E(\tau_0)},\label{L2-bound}
\end{eqnarray}
where in the last step we used~\eqref{energy-inequality_0}. Combining~\eqref{energy-inequality_0} with~\eqref{L2-bound} yields, for $\tau \geq \tau_0$,
\begin{align}
\|\phi(\tau,\,\cdot\,)&\|_{H^1(B_\tau)} \leq \nonumber \\
& C(1+\tau-\tau_0) \left(\|\phi(\tau_0,\,\cdot\,)\|_{H^1(B_{\tau_0})} + \|\partial_\tau\phi(\tau_0,\,\cdot\,)\|_{L^2(B_{\tau_0})}\right). \label{final_bound}
\end{align}
This estimate will be crucial in proving our main result.

\begin{proof}[{\bf Proof of Theorem~\ref{mainthm1}}]
Since $\lim_{t \to 0^+}a(t)=0$ and $1/{a(t)}$ is integrable in a right neighborhood of the origin, we can set $t_0=0$ in definition \eqref{tau}, thus obtaining $\lim_{t \to 0^+}\tau=0$ and $\lim_{\tau \to 0^+}a(\tau)=0$.
Note that conditions~\eqref{dado-inicial-posicao_0} and \eqref{dado-inicial-velocidade_0} in Theorem~\ref{mainthm1} can be written in terms of $\tau$ as
\begin{equation}\label{dado-inicial-posicao}
\lim_{\tau \to 0^+} \| \phi(\tau,\,\cdot\,) - A(\,\cdot\,) \|_{H^1(\Sigma)} = 0
\end{equation}
and
\begin{equation}\label{dado-inicial-velocidade}
\lim_{\tau \to 0^+} \|\partial_\tau \phi(\tau,\,\cdot\,) \|_{L^2(\Sigma)} = 0.
\end{equation}

Let $\tau_0>0$, and consider the Cauchy problem
\begin{equation}\label{ivp}
\begin{cases}
\Box_g \phi = 0, \\
\phi(\tau_0,\,\cdot\,) = A(\,\cdot\,), \\
\partial_\tau\phi(\tau_0,\,\cdot\,) = 0,
\end{cases} 
\end{equation}
where we assume for the time being that $A\in C^\infty(\Sigma) \cap H^3(\Sigma)$. As is well known (see for instance \cite{Sogge}), this problem has a unique solution $\phi^{\tau_0} \in C^\infty(M)$.

Fix $\tau_*\in\left(0,1\right)$ and assume that $0 < \tau_0 \leq \tau_1 \leq \tau_*$. 
Integrating~\eqref{waveqntau}, we obtain
\begin{equation}\label{dtauphi_bis}
\partial_\tau \phi^{\tau_0} (\tau_1,\,\cdot\,) = \frac{1}{a^{n-1}(\tau_1)} \int_{\tau_0}^{\tau_1} a^{n-1}(\tau)\Delta \phi^{\tau_0}(\tau,\,\cdot\,)\, d\tau.
\end{equation}
If $K \in \mathfrak{X}(\Sigma)$ is a Killing vector field of $h$, then $K \cdot \phi^{\tau_0}$ is again a solution of the wave equation with zero time derivative at $\tau=\tau_0$. 
Note that (see Appendix~\ref{appendixC}) there exist Killing vector fields $K_1, \ldots, K_N$ for the metric $h$ and integers $p_1, \ldots, p_N \in \{0,1\}$ such that
\begin{equation} \label{Laplacian}
\Delta\phi= \sum_{i=1}^N (-1)^{p_i} K_i \cdot (K_i \cdot \phi).
\end{equation}
From~\eqref{final_bound} we then have
\begin{equation}
\|\Delta \phi^{\tau_0}(\tau,\,\cdot\,)\|_{H^1(B_\tau)} \leq C\|A\|_{H^3(B)} \label{notthisone}
\end{equation}
for $\tau_0 \leq \tau \leq \tau_1$,
where $C$ denotes a universal constant (because $\tau_* < 1$). Using the fact that $a(\tau)$ is increasing, \eqref{dtauphi_bis} then yields 
\begin{equation}\label{velocity}
\|\partial_\tau \phi^{\tau_0} (\tau_1,\,\cdot\,)\|_{H^1(B_{\tau_1})} \leq 
C\|A\|_{H^3(B)} \, {\tau_1}.
\end{equation}
Since
$$
\phi^{\tau_0} (\tau_1,\,\cdot\,) = A(\,\cdot\,) + \int_{\tau_0}^{\tau_1} \partial_\tau\phi^{\tau_0}(\tau,\,\cdot\,)\, d\tau,
$$
we also obtain
\begin{equation}\label{position}
\|\phi^{\tau_0} (\tau_1,\,\cdot\,) - A(\,\cdot\,)\|_{H^1(B_{\tau_1})} \leq 
C\|A\|_{H^3(B)}\, {\tau_1}^2 .
\end{equation}

We may write~\eqref{velocity} and \eqref{position} (in reverse order) as
$$
\|\phi^{\tau_0} (\tau_1,\,\cdot\,) - 
\phi^{\tau_1} (\tau_1,\,\cdot\,)\|_{H^1(B_{\tau_1})} \leq 
C\|A\|_{H^3(B)}\, {\tau_1}^2
$$
and
$$
\|\partial_\tau \phi^{\tau_0} (\tau_1,\,\cdot\,)-
\partial_\tau \phi^{\tau_1} (\tau_1,\,\cdot\,)\|_{H^1(B_{\tau_1})} \leq 
C\|A\|_{H^3(B)} \, {\tau_1}.
$$

Since $\phi^{\tau_0} -\phi^{\tau_1}$ is a solution of the wave equation for
$\tau\geq\tau_1$, the energy inequality~\eqref{energy-inequality_0} and the a priori bound~\eqref{final_bound}
imply
$$
\|\phi^{\tau_0} (\tau,\,\cdot\,) - 
\phi^{\tau_1} (\tau,\,\cdot\,)\|_{H^1(B_\tau)} \leq 
C\|A\|_{H^3(B)}\, {\tau_1} \left( 1 + \tau \right)
$$
and
$$
\|\partial_\tau \phi^{\tau_0} (\tau,\,\cdot\,)-
\partial_\tau \phi^{\tau_1} (\tau,\,\cdot\,)\|_{L^2(B_\tau)} \leq 
C\|A\|_{H^3(B)} \, {\tau_1},
$$
for $\tau\geq\tau_1$  (recall that $\tau_1 \leq \tau_* < 1$). 

Since $A\in C^\infty(\Sigma) \cap H^3(\Sigma)$, all the previous estimates can be extended from geodesic balls to $\Sigma$. Moreover, as is well known (see for instance \cite{Sogge}), it suffices to take $A\in H^1(\Sigma)$ for the the initial value problem~\eqref{ivp} to have a unique (weak) solution 
\begin{equation} \label{solutionspace}
\phi^{\tau_0} \in C^0\left([\tau_0,\mathcal{T}\,],H^{1}(\Sigma)\right)\cap C^1\left([\tau_0,\mathcal{T}\,],L^2(\Sigma)\right),
\end{equation}
depending continuously on $A$ for these norms.\footnote{In fact, this result can be easily obtained from the estimates above by a standard approximation argument. The maximum conformal time $\mathcal{T}$ is related to the maximum time $T$ in the statement of the theorem by $\mathcal{T}=\int_0^T \frac{dt}{a(t)}$.} Since $C^\infty(\Sigma) \cap H^3(\Sigma)$ is dense in $H^3(\Sigma)$, and the injection $H^3(\Sigma) \subset H^1(\Sigma)$ is continuous, we conclude that for each $A\in H^3(\Sigma)$ there exists a unique (weak) solution of~\eqref{ivp} satisfying \eqref{solutionspace}, and that moreover this solution also satisfies all the estimates above for $\phi^{\tau_0}$, except~\eqref{notthisone}, extended to $\Sigma$.

For any $A\in H^3(\Sigma)$, define a sequence 
$(\phi_n)$ by 
$$
\phi_n=\phi^{\tau_n}\ \mbox{with}\ \tau_n=\frac{\tau_*}{n}.
$$
Given $\epsilon\in\left(0,\mathcal{T}\right)$,
it follows that $(\phi_n)_{\{n\geq N\}}$ is, for $N$ large enough, a Cauchy sequence 
of solutions of the wave equation in the Banach space
$$
C^0\left([\epsilon,\mathcal{T}\,],H^{1}(\Sigma)\right)\cap C^1\left([\epsilon,\mathcal{T}\,],L^2(\Sigma)\right).
$$
The limit, $\phi$, is a weak solution of the wave equation in this space. Since $\epsilon$ is arbitrary, we obtain a weak solution $\phi$ of the wave equation which is continuous in $(0,\mathcal{T}\,]$ with values in 
$H^{1}(\Sigma)$, and whose time derivative is continuous in $(0,\mathcal{T}\,]$ with values in $L^2(\Sigma)$.
 
Fix $\tau>0$. Going back to~\eqref{velocity} and~\eqref{position}, extended to $\Sigma$, we obtain
$$
\|\phi^{\tau_n} (\tau,\,\cdot\,) - A(\,\cdot\,)\|_{H^1(\Sigma)} \leq 
C\|A\|_{H^3(\Sigma)}\, {\tau}^2,
$$
$$
\|\partial_\tau \phi^{\tau_n} (\tau,\,\cdot\,)\|_{L^2(\Sigma)} \leq 
C\|A\|_{H^3(\Sigma)} \, {\tau},
$$
for $\tau\geq\tau_n$. Letting $n$ tend to $+\infty$, we obtain
$$
\|\phi (\tau,\,\cdot\,) - A(\,\cdot\,)\|_{H^1(\Sigma)} \leq 
C\|A\|_{H^3(\Sigma)}\, {\tau}^2,
$$
$$
\|\partial_\tau \phi (\tau,\,\cdot\,)\|_{L^2(\Sigma)} \leq 
C\|A\|_{H^3(\Sigma)} \, {\tau},
$$
for any $\tau > 0$. This proves~\eqref{dado-inicial-posicao} and \eqref{dado-inicial-velocidade}, and we can now guarantee that
$$
\phi \in C^0\left((0,\mathcal{T}\,],H^{1}(\Sigma)\right)\cap C^1\left((0,\mathcal{T}\,],L^2(\Sigma)\right).
$$ 

To prove uniqueness, we note that by using initial data in $C^\infty(\Sigma)$ approaching $\phi(\tau_0,\,\cdot\,)$ in $H^1(\Sigma)$ and $\partial_\tau\phi(\tau_0,\,\cdot\,)$ in $L^2(\Sigma)$, it is easy to extend the energy inequality~\eqref{energy-inequality_0} to weak solutions in
$$
C^0\left((0,\mathcal{T}\,],H^1(\Sigma)\right)\cap C^1\left((0,\mathcal{T}\,],L^2(\Sigma)\right)
$$
with $0<\tau_0<\tau_1\leq \mathcal{T}$. Taking the limit of this inequality as $\tau_0 \to 0^+$ yields uniqueness. 
\end{proof}
%
%
\section*{Acknowledgements}
We thank Artur Alho and Anne Franzen for valuable comments and suggestions. We especially thank Hans Ringstr\"om for his detailed remarks concerning the relation between his results and Theorem~\ref{mainthm1}. This work was partially supported by FCT/Portugal through UID/MAT/04459/2013 and grant (GPSEinstein) PTDC/MAT-ANA/1275/2014.
%
%
\appendix
%
%
\section{FLRW models in $n+1$ dimensions}\label{appendixA}
In this appendix we present some examples of FLRW models satisfying the hypotheses of Theorem~\ref{mainthm1}. We assume that the FLRW metric
$$
g = -dt^2 + a^2(t) h_{ij} dx^i dx^j 
$$
solves the $(n+1)$-dimensional Einstein field equations
$$
R_{\mu\nu} - \frac12 R g_{\mu\nu} + \Lambda g_{\mu\nu} = \kappa T_{\mu\nu}
$$
where $\kappa > 0$ is the $(n+1)$-dimensional gravitational coupling constant. If we take the energy-momentum tensor to be that of a comoving perfect fluid with a linear equation of state,
$$
T = (\rho + P) dt^2 + P g, \qquad P = (\gamma - 1) \rho,
$$
then the Einstein equations become equivalent to the Friedmann equations
$$
\rho = \rho_0 a^{-n\gamma}
$$
and
$$
\dot{a}^2 = \frac{2\kappa\rho_0}{n(n-1)} a^{-n\gamma + 2} + \frac{2\Lambda}{n(n-1)} a^2 - k,
$$
where $k=0, 1$ or $-1$ is the spatial curvature (see~\cite{ChenGibbonsLiYang}). It is easy to see that if we take
$$
\gamma > \frac2n
$$
then these equations always lead to a Big Bang, and we have
$$
a(t) \sim t^p, \qquad p = \frac2{n\gamma}
$$
as $t \to 0^+$ (for an appropriate choice of units). In fact, this form of $a(t)$ is actually an exact solution when $\Lambda = k = 0$. Note that the condition on $\gamma$ above is equivalent to
$$
p \in \left(0,1\right),
$$
so that the hypotheses of Theorem~\ref{mainthm1} are satisfied.

The fluid's speed of sound is given by
$$
c_s = \sqrt{\frac{dP}{d\rho}} = \sqrt{\gamma - 1}.
$$
Important special cases are $\gamma=1$ (dust), $\gamma=1+\frac1n$ (radiation) and $\gamma=2$ (stiff fluid). Note that $\gamma > 2$, that is, $p < \frac1n$, corresponds to a speed of sound larger than the speed of light, and is not regarded as physical. 
%
%
\section{Killing vector fields and the Laplacian}\label{appendixC}
This appendix contains a brief proof that~\eqref{Laplacian} holds in $\bbR^n$, $\mathbb{S}^n$ and $\mathbb{H}^n$. 

The case when $(\Sigma,h)$ is $\bbR^n$ with the Euclidean metric is trivial, since it suffices to take $N=n$, $p_i=0$ and $K_i = \partial_i$.

When $(\Sigma,h)$ is $S^n$ with the unit round metric, we can compute $\Delta\phi$ by extending $\phi$ to a neighborhood of $S^n$ in $\bbR^{n+1}$ as a radially constant function, and restricting the Laplacian of this extension to $S^n$. The same is true for the hyperbolic space $\mathbb{H}^n$, seen as the set of future-pointing unit timelike vectors in $\bbR^{n+1}$ with the Minkowski metric. 
In both cases, let $g$ be either the Euclidean or the Minkowski metric on $\bbR^{n+1}$, and consider the linear Killing vector fields
$$
K^{\alpha\beta} = \left( g^{\alpha\mu} g^\beta_\nu - g^{\beta\mu} g^\alpha_\nu \right) x^\nu \partial_\mu.
$$
These fields are tangent to either $\mathbb{S}^n$ or $\mathbb{H}^n$, since these hypersurfaces correspond to unit vectors and the isometries generated by linear Killing vector fields fix the origin. Therefore, they can be interpreted as Killing vector fields on either $\mathbb{S}^n$ or $\mathbb{H}^n$. 
If $\phi$ is a radially constant function,
$$
x^\alpha \partial_\alpha \phi = 0,
$$
then
\begin{align*}
K^{\alpha\beta} \cdot K_{\alpha\beta} \cdot \phi & = \left( g^{\alpha\mu} g^\beta_\nu - g^{\beta\mu} g^\alpha_\nu \right) x^\nu \partial_\mu \left[\left( g_\alpha^\rho g_{\beta\sigma} - g_\beta^\rho g_{\alpha\sigma} \right) x^\sigma \partial_\rho \phi \right] \\
& = \left( 2g^\rho_\nu - 2(n+1) g^\rho_\nu \right) x^\nu \partial_\rho \phi + \left( 2g^{\rho\mu} g_{\sigma\nu} - 2 g^\mu_\sigma g^\rho_\nu \right) x^\nu x^\sigma \partial_\mu \partial_\rho \phi \\
& = \left(2 g_{\sigma\nu} x^\nu x^\sigma\right) g^{\rho\mu} \partial_\mu \partial_\rho \phi,
\end{align*}
where we used
$$
x^\mu x^\rho \partial_\mu \partial_\rho \phi = x^\mu \partial_\mu \left( x^\rho \partial_\rho \phi \right) - x^\mu \partial_\mu\phi = 0.
$$

In the Euclidean case, restricting to the sphere $S^n \subset \bbR^{n+1}$, we have
$$
K^{\alpha\beta} \cdot K_{\alpha\beta} \cdot \phi = 2 \Delta \phi,
$$
or, equivalently,
$$
\sum_{\alpha < \beta} K^{\alpha\beta} \cdot K_{\alpha\beta} \cdot \phi = \Delta \phi.
$$
Hence, in this case we can take the $N=\frac{n(n+1)}2$ Killing vector fields $K_{\alpha\beta}$ for $\alpha < \beta$ and $p_1 = \ldots = p_N =0$ to obtain \eqref{Laplacian}.

In the Minkowski case, restricting to the hypersurface $\mathbb{H}^n \subset \bbR^{n+1}$ formed by the future-pointing unit timelike vectors, we have
$$
K^{\alpha\beta} \cdot K_{\alpha\beta} \cdot \phi = - 2 \Box \phi,
$$
or, equivalently,
$$
\sum_{\alpha < \beta} - K^{\alpha\beta} \cdot K_{\alpha\beta} \cdot \phi = \Box \phi.
$$
Hence, in this case we can again take the $N=\frac{n(n+1)}2$ Killing vector fields $K_{\alpha\beta}$ for $\alpha < \beta$, but now with $p_1=\ldots=p_n=0$ (corresponding to $K_{01}, \ldots, K_{0n}$) and $p_{n+1}=\ldots=p_N=1$ (corresponding to $K_{ij}$ with $1 \leq i < j$), to obtain \eqref{Laplacian}.
%
%

\end{document}